\documentclass[12pt]{article}
\usepackage{epsfig, epsf, graphicx, subfigure}
\usepackage{pstricks, pst-node, psfrag}
\usepackage{amssymb,amsmath,bm}
\usepackage{verbatim,enumerate}
\usepackage{rotating, lscape}
\usepackage{setspace}
\usepackage[english]{babel}
\usepackage[sort&compress]{natbib}
\usepackage{multirow}
\usepackage{theorem}
\usepackage{floatrow}
\usepackage{caption}
\usepackage{float}
\floatstyle{plaintop}
\restylefloat{table}

\def\Cov{\mathrm{cov}}
\def\Cor{\mathrm{cor}}

\def\d{\mathrm{d}}

\def\p{\partial}

\def\00{\mathrm{0}}

\def\Uu{\mathbf{U}}
\def\Ww{\boldsymbol{W}}
\def\Zz{\mathbf{Z}}

\def\WW{\mathrm{W}}
\def\tht{\boldsymbol{\theta}}
\def\alp{\boldsymbol{\alpha}}

\def\ee{\mathcal{E}}
\def\pp{\mathcal{P}}
\def\ss{\boldsymbol{s}}

\newtheorem{prop}{Proposition}

\begin{document}

\thispagestyle{empty} \baselineskip=28pt \vskip 5mm
\begin{center} {\Huge{\bf A Copula Model for Non-Gaussian Multivariate Spatial Data}}
\end{center}

\baselineskip=12pt \vskip 10mm

\begin{center}\large
Pavel Krupskii and 
Marc G. Genton\footnote[1]{
\baselineskip=10pt Statistics Program,
King Abdullah University of Science and Technology,
Thuwal 23955-6900, Saudi Arabia. E-mail: pavel.krupskiy@kaust.edu.sa, 
marc.genton@kaust.edu.sa\\
This research was supported by the
King Abdullah University of Science and Technology (KAUST).}
\end{center}

\baselineskip=17pt \vskip 10mm \centerline{\today} \vskip 15mm

\begin{center}
{\large{\bf Abstract}}
\end{center}

We propose a new copula model for replicated multivariate spatial data. Unlike classical models that assume multivariate normality of the data, the proposed copula is based on the assumption that some factors exist that affect the joint spatial dependence of all measurements of each variable as well as the joint dependence among these variables. The model is parameterized in terms of a cross-covariance function that may be chosen from the many models proposed in the literature. In addition, there are additive factors in the model that allow tail dependence and reflection asymmetry of each variable measured at different locations and of different variables to be modeled. The proposed approach can therefore be seen as an extension of the linear model of coregionalization widely used for modeling multivariate spatial data. 
The likelihood of the model can be obtained in a simple form and therefore the likelihood estimation is quite fast. 
The model is not restricted to the set of data 
locations, and using the estimated copula, spatial data can be interpolated at locations where values of variables are unknown. We apply the proposed model to temperature and pressure data and compare its performance with the performance of a popular model from multivariate geostatistics.

\baselineskip=14pt

\par\vfill\noindent
{\bf Some key words:}
Copula; Heavy tails; Permutation asymmetry; Spatial statistics; Tail asymmetry.
\par\medskip\noindent
{\bf Short title}: A Copula Model for Non-Gaussian Multivariate Spatial Data

\clearpage\pagebreak\newpage \pagenumbering{arabic}
\baselineskip=26pt

\section{Introduction}

Modeling spatial data is a challenging task as it often requires flexible, but simple and tractable models that can handle multivariate data. 
In many applications, models for multivariate spatial data are of interest for co-kriging; see \cite{Furrer.Genton2011} and references therein. To set up the problem, we assume that we have a $p$-dimensional random process, $\mathbf{Z}(\ss) = \{Z_1(\ss),\ldots,Z_p(\ss)\}^{\top}$, defined at locations $\ss \in \mathbb{R}^d$. In this case, several variables are observed at different locations and the main task is to model dependence both within each variable, $Z_i(\ss)$, $i=1,\ldots,p$ (at different locations), and between different variables. If $Z(\ss)$ is a Gaussian multivariate second-order stationary random process with zero mean, only a covariance structure of $Z(\ss)$ is needed to completely characterize the dependence structure of the process; i.e., the cross-covariance functions $C_{i_1,i_2}(\ss_1,\ss_2)=\Cov\{Z_{i_1}(\ss_1),Z_{i_2}(\ss_2)\}$ need to be modeled. A popular approach is to model a matrix of correlations for each variable and cross-correlations of different variables using a multivariate covariogram or a cross-variogram \citep{Wackernagel2003, Chiles.Delfiner1999} or using pseudo cross-variograms \citep{Myers1991}. The linear model of coregionalization (LMC) has been widely used to model cross-covariances; see \cite{Goulard.Voltz1992}, \cite{Wackernagel2003}, \cite{Gelfand.Schmidt.ea2004}, \cite{Apanasovich.Genton.2010}, and the review by \cite{Genton.Kleiber2015} and references therein.

The LMC for a stationary Gaussian process with zero mean assumes a linear structure:
$$Z_i(\ss) = \sum_{k=1}^rA_{ik}\tilde Z_k(\ss),$$
where we assume that $\tilde Z_1,\ldots,\tilde Z_r$ are independent stationary Gaussian processes to keep computations simple,
so that $$C_{i_1,i_2}(\ss_1,\ss_2) = C_{i_1,i_2}(\ss_1 - \ss_2) = \sum_{k=1}^rC_k(\ss_{i_1}-\ss_{i_2})A_{i_1k}A_{i_2k}, \quad 1 \leq r \leq p,$$ where $A$ is a $p \times r$ full rank matrix and $C_k(\cdot)$ are valid stationary covariance functions. Different structures for the covariance functions have been proposed in the literature, such as the Mat\'{e}rn covariance function; see for example \cite{Gneiting2002} and \cite{Gneiting.Genton.ea2007} for a review of covariance functions. Parameters in the LMC can be estimated using the least squares method \citep{Goulard.Voltz1992} or using maximum likelihood \citep{Zhang2007}.

Despite its simplicity and tractability, the LMC focuses on modeling the cross-covariance structure and may not be appropriate if the joint normality assumption is not valid for the multivariate spatial data under consideration. This can happen, for example, with data with strong joint dependence in the tails (i.e., when large/small values are simultaneously observed more often in repeated measurements than predicted by the model), or with data with reflection asymmetry (i.e., when large values are simultaneously observed more often in repeated measurements than small values, or vice versa).

To overcome this problem, more flexible copula-based models can be considered. Copulas have been used in a wide range of actuarial,
financial and environmental studies; see \cite{Krupskii.Joe2015b}, \cite{Genest.Favre2007}, \cite{Patton2006} and others. A copula is a multivariate cumulative distribution function with univariate uniform $U(0,1)$ marginals; this function can be used to link univariate marginals to construct the joint distribution function. \cite{Sklar1959} showed that for a continuous $n$-dimensional distribution function, $F$, and its univariate marginal distribution functions, $F_1,\ldots,F_n$, there exists a unique copula function, $C$, such that $F(z_1,\ldots,z_n) = C\{F_1(z_1),\ldots,F_n(z_n)\}$ for any $z_1,\ldots,z_n$.


Some copula-based models have been proposed in the literature to model univariate spatial processes. One popular approach is to use vine copula models in which the joint distribution is constructed using bivariate linking copulas and from which different types of dependence structures can be obtained; see \cite{Kurowicka.Cooke2006} and \cite{Aas.Czado.ea2009} for details. The models using vine copulas have been applied to study climate, geology, radiation and other spatial data; see \cite{Graler.Pebesma2011}, \cite{Graler2014}, \cite{Erhardt.Czado.ea2014} and others. In these models, the dependence structure is selected based on the likelihood, making interpretability difficult. Moreover, likelihood estimation can be quite slow in high dimensions.

Other copula models for univariate spatial data include copulas parameterized in terms of pairwise dependencies, for instance a v-transformed copula of \cite{Bardossy.Li2008} and a chi-squared copula of \cite{Bardossy2006}. These copulas are constructed by using a non-monotonic transformation of multivariate normal variables. As such, they cannot be used for modeling tail dependence. Moreover, the likelihood function has no simple form and obtaining parameter estimates in these models is a difficult task. Recently, \cite{Krupskii.Huser.ea2016} proposed a factor copula model for spatial data that allows tail dependence and reflection asymmetry to be modeled. Likelihood estimates in that model can be obtained quite easily with some choice of the common factor, even if the number of locations is fairly large. However, to the best of our knowledge, flexible copula models have not yet been studied for multivariate spatial data with replicates, i.e., when there are several variables and spatial locations and each variable is repeatedly measured at each spatial location.


In this paper, we extend the approach of \cite{Krupskii.Huser.ea2016} and propose a model for stationary spatial processes that combines the flexibility of a copula modeling approach, the interpretability of the LMC, and the tractability of the normal copula in high dimensions. The model and the corresponding copula are based on the following multivariate random process:
\begin{equation}
\label{base_model}
W_i(\ss) = Z_i(\ss) + \alpha_{i0}^U\ee_0^U + \alpha_{i}^U\ee_{i}^U - \alpha_{i0}^L\ee_0^L - \alpha_{i}^L\ee_i^L \quad (s \in \mathbb{R}^d, \, i =1,\ldots,p),
\end{equation}
where $Z_i(\ss)$ are cross-correlated Gaussian processes, $\alpha_{i0}^U, \alpha_i^U, \alpha_{i0}^L, \alpha_i^L \geq 0$ for identifiability and $\ee_{i}^U,\ee_{i}^L,\ee_0^U,\ee_0^L \sim \mathrm{Exp}(1)$ are independent common factors with unit exponential distribution that do not depend on the spatial location, $s$. This model can therefore be suitable for modeling spatial data in a small domain when some common factors affect the joint dependence of the variables. One example is data collected by weather stations from a small region subject to common weather conditions.

The choice of the common factors allows different types of dependence structures to be generated and makes parameter estimation in this model fairly easy. The joint dependence of the Gaussian processes, $Z_i(\ss)$, can be modeled using the LMC. The proposed model uses additive independent exponential factors to introduce tail dependence and tail asymmetry  and therefore it can be seen as an extension of the LMC model where the cross-covariance function can be constructed using a sum of independent processes. Of course, different cross-covariance functions for $Z_i(\ss)$ can be considered as well; see the review by \cite{Genton.Kleiber2015}.


In our model, as well as in many other copula-based models, replicates are needed to estimate dependence parameters. With different sets of parameters in (\ref{base_model}) models with the same covariance and cross-covariance structures but with different tail properties can be obtained. Repeated measurements of the multivariate spatial process are thus needed to estimate dependence both in the middle of the joint distribution and in its tails. The proposed model is therefore suitable for modeling processes that can be repeatedly measured in time. These include weather data (temperature, atmospheric pressure, wind speed and direction), pollution levels, and satellite data, to name but a few. Measurements can be correlated in time and therefore some autoregressive models can be applied to fit the marginal distributions (at fixed locations) and the joint distribution of the residuals (across all locations) can be modeled using the proposed copula model. In other words, temporal dependence in spatio-temporal multivariate data can be removed by fitting appropriate marginal models to each variable.  The vectors of residuals can then be treated as replicates if there is no significant dependence between residuals for different variables at different time lags.


In this paper, we focus on the bivariate case, $p=2$, but we also show how our approach can be extended to $p>2$. The remainder of the paper is organized as follows. In Section \ref{sec_factmodel}, we describe the model (\ref{base_model}) in detail and study its dependence properties. We first define the model in a general case with $p \geq 2$, and then, in the following sections, we provide more details about the bivariate case,  $p=2$. 
More details on the likelihood estimation, assessing goodness of fit and interpolation of a spatial process at a new location
are given in Section \ref{sec_MLE}. 
We apply the proposed copula model to bivariate spatial data of temperature and atmospheric pressure in Oklahoma, USA, in Section \ref{sec_empstudy}, and we conclude with a discussion in Section \ref{sec_discussion}.

\section{The model}
\label{sec_factmodel}

We use the following notation:
$\Phi(\cdot)$ is the cumulative distribution function of the univariate standard normal random variable, whereas $\Phi_{\Sigma}(\cdot)$ is that of the multivariate standard normal random vector with correlation matrix $\Sigma$. For simplicity, in the bivariate case with $\Sigma_{12} = \rho$, we use the notation $\Phi_{\rho}(\cdot)$. Small symbols denote the corresponding densities.

We consider measurements of a random multivariate process by assuming that unobserved random factors exist that affect the joint dependence of all measurements of each variable as well as the joint dependence between every two variables. Specifically, we construct the corresponding copula by restricting model (\ref{base_model}) to a finite set of locations $s_1,\ldots,s_n \in \mathbb{R}^d$. Let $\{(W_{1j},\ldots,W_{pj})\}_{j=1}^n$ be measurements of a $p$-variate spatial process that is observed at $n$ different locations and let
\begin{equation}
\label{mainmodel}
W_{ij} = Z_{ij} + \alpha_{i0}^U\ee_0^U + \alpha_{i}^U\ee_{i}^U - \alpha_{i0}^L\ee_0^L - \alpha_{i}^L\ee_i^L \quad (j=1,\ldots,n, \, i =1,\ldots,p),
\end{equation}
where $\ee_i^U,\ee_i^L,\ee_0^U,\ee_0^L \sim_{\mathrm{i.i.d.}} \mathrm{Exp}(1)$ are exponential common factors that are independent of $Z_{ij}$ and where  $\Zz=(Z_{11},\ldots,Z_{1n},\ldots,Z_{p1},\ldots,Z_{pn})^{\top}$ has a multivariate normal distribution with standard normal marginals and some covariance matrix, $\Sigma_{\Zz}$. Exponential factors allow flexible dependence structures to be generated such that parameter estimation becomes quite fast. 
The structure of $\Sigma_{\Zz}$ depends on the model for $Z$; for example, one may use LMC. The correlation structure of $W$ depends on that of $Z$. For the $i$-th variable, $$\Cor(W_{i,j_1},W_{i,j_2})=\frac{\Cor(Z_{i,j_1},Z_{i,j_2})+(\alpha_{i0}^U)^2+(\alpha_{i}^U)^2+(\alpha_{i0}^L)^2+(\alpha_{i}^L)^2}{1+(\alpha_{i0}^U)^2+(\alpha_{i}^U)^2+(\alpha_{i0}^L)^2+(\alpha_{i}^L)^2}\,,$$ and for the $i_1$-th and $i_2$-th variables ($i_1 \neq i_2$), $$\Cor(W_{i_1,j_1},W_{i_2,j_2})=\frac{\Cor(Z_{i_1,j_1},Z_{i_2,j_2})+\alpha_{i_10}^U\alpha_{i_20}^U+\alpha_{i_20}^L\alpha_{i_20}^L}{[\{(\alpha_{i_10}^U)^2+(\alpha_{i_10}^L)^2\}\{(\alpha_{i_20}^U)^2+(\alpha_{i_20}^L)^2\}]^{1/2}}\,.$$
Note that $\Cov(Z_{i,j_1},Z_{i,j_2})=1$ implies $\Cov(W_{i,j_1},W_{i,j_2})=1$; this corresponds to perfect co-monotonic dependence.

Let $\Ww = (W_{11},\ldots,W_{1n},\cdots,W_{p1},\ldots,W_{pn})^{\top}$ and let $F_{n,p}^{\Ww}$ and $f_{n,p}^{\Ww}$ respectively be the cumulative distribution function and probability density function of the vector $\Ww$. The function $f_{n,p}^{\Ww}$ can be obtained in a simple form; we provide more details for $p=2$ in Appendix \ref{appx-pdf}. Let $F_{1,i}^{W}$ and $f_{1,i}^{W}$ respectively be the cumulative distribution function and probability density function of 
$W_{i1}$ ($i=1,\ldots,p$). Let $$\xi(z;\alpha_i^L,\alpha_i^U,\alpha_{i0}^L,\alpha_{i0}^U) = \frac{(\alpha_{i}^U)^3\exp\left\{0.5/(\alpha_{i}^U)^2-z/\alpha_{i}^U\right\}\Phi(z-1/\alpha_i^U)}{\{(\alpha_{i0}^L+\alpha_{i}^U)(\alpha_{i}^L+\alpha_{i}^U)(\alpha_{i0}^U-\alpha_{i}^U)\}}\,.$$ One can show that
\begin{eqnarray*}
F_{1,i}^{W}(z) &=& \Phi(z)+\xi(z;\alpha_i^L,\alpha_i^U,\alpha_{i0}^L,\alpha_{i0}^U)
-\xi(-z;\alpha_i^U,\alpha_i^L,\alpha_{i0}^U,\alpha_{i0}^L)\\
&& +\,\xi(z;\alpha_{i0}^L,\alpha_{i0}^U,\alpha_{i}^L,\alpha_{i}^U)
-\xi(-z;\alpha_{i0}^U,\alpha_{i0}^L,\alpha_{i}^U,\alpha_{i}^L).
\end{eqnarray*}
Because $F_{1,i}^{W}(z)$ takes a simple form, the inverse function, $(F_{1,i}^{W})^{-1}(z)$, can be easily calculated using numerical methods. Let $\mathbf{u}_i=(u_{i1},\ldots,u_{in})^{\top}$, $0 \leq u_{ij} \leq 1$, $j=1,\ldots,n$. The copula and its density corresponding to the distribution of $W$ ($C_{n,p}^{\Ww}$ and $c_{n,p}^{\Ww}$\,, respectively) can then be obtained as follows:
\begin{eqnarray}
C_{n,p}^{\Ww}(\mathbf{u}_1,\ldots,\mathbf{u}_p) &=& F_{n,p}^{\Ww}\left\{(F_{1,1}^{W})^{-1}(\mathbf{u}_1),\ldots,(F_{1,p}^{W})^{-1}(\mathbf{u}_p)\right\}, \nonumber\\
c_{n,p}^{\Ww}(\mathbf{u}_1,\ldots,\mathbf{u}_p) &=& \frac{f_{n,p}^{\Ww}\left\{(F_{1,1}^{W})^{-1}(\mathbf{u}_1),\ldots,(F_{1,p}^{W})^{-1}(\mathbf{u}_p)\right\}}{f_{1,1}^{W}\left\{(F_{1,1}^{W})^{-1}(\mathbf{u}_1)\right\}\times\cdots\times f_{1,p}^{\Ww}\left\{(F_{1,p}^{W})^{-1}(\mathbf{u}_p)\right\}}\,. \label{cop_pdf}
\end{eqnarray}
Here, the marginal distributions $F^{W}_{1,i}$, $i=1,\ldots,p$, need not be the marginal distributions of the original data; these distributions are only used to construct the joint copula cumulative distribution function. The copula $C^{\Ww}_{n,p}$ can be combined with arbitrary univariate marginals, depending on the data, thus allowing greater flexibility in the proposed model.

Spatial data often have strong dependence in the tails and therefore a model that can handle strong tail dependence is necessary. One standard approach to measure tail dependence for a bivariate copula, $C$, is to use the lower and upper tail dependence coefficients, $\lambda_L$ and $\lambda_U$, respectively:
$$
\lambda_L = \underset{q\to 0}{\lim}\,C(q,q)/q \in [0,1] \quad \text{and} \quad \lambda_U = \underset{q\to 0}{\lim}\,\bar C(1-q,1-q)/q \in [0,1],
$$
where $\bar C(u_1,u_2)=1-u_1-u_2+C(u_1,u_2)$ is the survival copula. The copula $C$ is said to have lower (upper) tail dependence if $\lambda_L > 0$ ($\lambda_U > 0$). For the  normal copula, $\lambda_L = \lambda_U = 0$. Models based on multivariate normality 
are therefore not suitable for modeling data with strong tail dependence.

Dependence properties of the copula $C_{n,p}^{\Ww}$ depend on the choice of the parameters, $\alpha_{i}^L,\alpha_{i}^U,\alpha_{i0}^L$, $\alpha_{i0}^U$ ($i=1,\ldots,p$). For the $i$-th variable, the proposed copula simplifies to the one introduced in \cite{Krupskii.Huser.ea2016}, with the common factor $V_0 = \alpha_{i0}^U\ee_0^U + \alpha_{i}^U\ee_{i}^U - \alpha_{i0}^L\ee_0^L - \alpha_{i}^L\ee_i^L$. In particular, it follows that the bivariate copula, $C_{2,i}^{\Ww}$, corresponding to the distribution of $(u_{i1},u_{i2})$, has lower and upper tail dependence with $\lambda_L = 2\Phi\left[-\{(1-\rho_{1,2}^i)/2\}^{1/2}/\tilde\alpha_i^L\right]$ and $\lambda_U = 2\Phi\left[-\{(1-\rho_{1,2}^i)/2\}^{1/2}/\tilde\alpha_i^U\right]$, where $\rho_{1,2}^i = \Cor(Z_{i1},Z_{i2})$, $\tilde\alpha_i^L = \max(\alpha_i^L, \alpha_{i0}^L)$ and $\tilde\alpha_i^U = \max(\alpha_i^U, \alpha_{i0}^U)$.

The exponential distribution for the common factor $V_0$ is selected for two reasons. First, 
if the distribution of $V_0$ has heavier tails, such as the Pareto factor, the tail dependence coefficients for $C_{2,i}^{\Ww}$ always equal one and therefore the dependence in the tails does not weaken with distance. At the same time, if the distribution of $V_0$ has lighter tails, the resulting copula does not have tail dependence \citep{Krupskii.Huser.ea2016}. Second, the likelihood function has a simple form for this choice of $V_0$ which makes parameter estimation much faster.

We now investigate dependence between two different variables. Without loss of generality, we consider the copula $C_{2,1:2}^{\Ww}$, corresponding to the distribution $F_{2,1:2}^{\Ww}$ of $(u_{11},u_{21})$ with $\rho_{1,2}^{1:2} := \Cov(Z_{11},Z_{21})$.

We define $\ell_n(x_1,x_2) := n[1-F_{2,1:2}^{\Ww}\{(F_{1,1}^{W})^{-1}(1-x_1/n),(F_{1,2}^{W})^{-1}(1-x_2/n)\}]$. The limit $\ell(x_1,x_2) := \lim_{n\to \infty} \ell_n(x_1,x_2)$ is called the stable upper tail dependence function of the limiting extreme value copula; see \cite{Segers.2012a}. We next show when the copula $C_{2,1:2}^{\Ww}$ has upper tail dependence and compute the limit $\ell(x_1,x_2)$. For simplicity, we assume that $\alpha_{10}^L = \alpha_{20}^L = \alpha_1^L = \alpha_2^L = 0$. A similar result holds in the general case and for the lower tail as well; however, in the case of tail dependence, the formula for $\ell(x_1,x_2)$ is more complicated when the coefficients $\alpha_{10}^L, \alpha_{20}^L, \alpha_1^L$ and $\alpha_2^L$ are nonzero.

\begin{prop}  \label{prop-1} \rm
Let $\delta_1 = \alpha_{10}^U/\alpha_{1}^U$, $\delta_2 = \alpha_{20}^U/\alpha_{2}^U$ and $\delta_{12} = \delta_1+\delta_2$. Denote by $y_i=x_i(1-1/\delta_i)$ and $\delta_i^* = (\delta_i-1)^{-1}-(\delta_{12}-1)^{-1}$ ($i=1,2$). If $\min(\delta_1, \delta_2) < 1$, the copula $C_{2,1:2}^{\Ww}$ has no upper tail dependence and $\ell(x_1,x_2)=x_1+x_2$. If $\min(\delta_1,\delta_2) > 1$, copula $C_{2,1:2}^{\Ww}$ has upper tail dependence and, with $\rho_{12}:=\{(\alpha_{10}^U)^2-2\rho_{1,2}^{1:2}\alpha_{10}^U\alpha_{20}^U+(\alpha_{20}^U)^2\}^{1/2}/(\alpha_{10}^U\alpha_{20}^U)$,
\begin{eqnarray}
\ell(x_1,x_2) &=& \frac{\delta_1y_1}{\delta_1-1}\Phi\left\{\frac{\rho_{12}}{2}+\frac{\log(y_1/y_2)}{\rho_{12}}\right\} + 
\frac{\delta_2y_2}{\delta_2-1}\Phi\left\{\frac{\rho_{12}}{2}+\frac{\log(y_2/y_1)}{\rho_{12}}\right\} \nonumber\\
&&+ y_2^{\delta_2}y_1^{1-\delta_2}\delta_2^*\exp\left\{0.5\delta_2(\delta_2-1)\rho_{12}^2\right\}\Phi\left\{\rho_{12}(0.5-\delta_2)+\frac{\log(y_1/y_2)}{\rho_{12}}\right\}
\nonumber \\
&&+ y_1^{\delta_1}y_2^{1-\delta_1}\delta_1^*\exp\left\{0.5\delta_1(\delta_1-1)\rho_{12}^2\right\}\Phi\left\{\rho_{12}(0.5-\delta_1)+\frac{\log(y_2/y_1)}{\rho_{12}}\right\}\,.
\label{prop-1-HS}
\end{eqnarray}
\end{prop}
The proof is given in Appendix \ref{appx-prop1}.

\emph{Remark 1}. The limiting extreme value copula corresponding to $C_{2,1:2}^{\Ww}$ is $\mathcal{C}_{2,1:2}^{\Ww}(u_1, u_2) = \exp\{-\ell(-\log u_1, -\log u_2)\}$. When $\alpha_{1}^U = \alpha_{2}^U = 0$ and $\min(\delta_1, \delta_2) > 1$, $\mathcal{C}_{2,1:2}^{\Ww}$ is the H\"{u}sler-Reiss copula with parameter $\lambda = \rho_{12}$; see \cite{Husler.Reiss1989} for more details on the H\"{u}sler-Reiss bivariate distribution. In the general case, $\mathcal{C}_{2,1:2}^{\Ww}$ is permutation symmetric if and only if $\delta_1 = \delta_2$; that is, it is permutation symmetric if $\alpha_{1}^U\alpha_{20}^U = \alpha_{2}^U\alpha_{10}^U$. Permutation asymmetry of the extreme-value copula can be useful in applications for modeling data that are permutation asymmetric in the tails. 

\emph{Remark 2}. The upper tail dependence coefficient for $C_{2,1:2}^{\Ww}$ is $2-\ell(1,1)$. Using the result of Proposition \ref{prop-1}, one can check that this is a monotonically increasing function of $\rho_{1,2}^{1:2}$. When two variables are measured at two different locations, $\rho_{1,2}^{1:2}$ is a decreasing function of the distance between these locations if parameterized using one of the many cross-covariance models proposed in the literature. This implies that the upper tail dependence of $C_{2,1:2}^{\Ww}$ is also a decreasing function of the distance. The strength of this dependence is controlled by the parameters $\delta_i$, $i=1,2$, to allow greater flexibility of the proposed copula model.
\section{Maximum Likelihood Estimation and Interpolation}
\label{sec_MLE}

\subsection{The likelihood function}
\label{sec_MLE_sub0}

We now show how to obtain the maximum likelihood estimates for the copula parameters in model (\ref{mainmodel}).
We assume that we observe $N$ independent samples, $\mathbf{w}_1,\ldots,\mathbf{w}_N$, from model (\ref{mainmodel}), where $\mathbf{w}_k=(\mathbf{w}_{1,k},\ldots,\mathbf{w}_{p,k})^{\top}$, $\mathbf{w}_{i,k} = (w_{i1,k},\ldots,w_{in,k})^{\top}$  ($i=1,\ldots,p$, $k=1,\ldots,N$) with essentially arbitrary marginals, not necessarily given by cumulative distribution functions $F_{1,i}^{W}$. Here, the vector $\mathbf{w}_{i,k}$ represents the $k$-th replicate of the $i$-th variable measured at $n$ different locations. To estimate the copula parameters, we need to transform the data to a uniform scale, e.g., non-parametrically, as follows: for each $i=1,\ldots,p$ and $j=1,\ldots,n$, we can define the uniform scores, $u_{ij,k} = \{\mathtt{rank}(w_{ij,k})-0.5\}/N$ ($k=1,\ldots,N$).
We let $\mathbf{z}_k=(\mathbf{z}_{1,k},\ldots,\mathbf{z}_{d,k})^{\top}$, $\mathbf{z}_{i,k} = (z_{i1,k},\ldots,z_{in,k})^{\top}$, $z_{ij,k} = (F_{1,i}^{W})^{-1}(u_{ij,k};\tht_{F,i})$, where $\tht_{F,i}$ is a vector of parameters for $F_{1,i}^{W}$  ($i=1,\ldots,p$,\,  $j=1,\ldots,n$,\, $k=1,\ldots,N$). Because we use data transformed to uniform scores, they are an approximation to $U(0,1)$ data. Therefore, the dependence parameters can be estimated via a pseudo-likelihood function. As the number of replicates goes to infinity, $N \to \infty$, the likelihood estimates are consistent and asymptotically normal provided that the copula is correctly specified; see chapter 5.9 of \cite{Joe2014} for details. Let $\tht_F=(\tht_{F,1},\ldots,\tht_{F,p})^{\top}$. From (\ref{cop_pdf}), the pseudo log-likelihood is:
\begin{equation}
l(\mathbf{z}_1,\ldots,\mathbf{z}_N) = \sum_{k=1}^N\log f_{n,p}^{\Ww}(\mathbf{z}_{1,k},\ldots,\mathbf{z}_{p,k};\tht_F,\tht_{\Sigma}) - \sum_{k=1}^N\sum_{i=1}^p\sum_{j=1}^n\log f_{1,i}^{W}(z_{ij,k};\tht_{F,i}), \label{loglik}
\end{equation}
where $\tht_{\Sigma}$ is a vector used to parameterize the correlation matrix, $\Sigma_{Z}$.

The full model for a $p$-variate spatial process has $4p$ tail parameters $\alpha_i^L, \alpha_{i0}^L$ and $\alpha_i^U, \alpha_{i0}^U$ to be estimated together with the parameters for the covariance matrix, $\tht_{\Sigma}$. Simulations show that it might be difficult to obtain accurate estimates when the sample size is not large ($N \leq 500$). The main reason is that different sets of these parameters may result in models with similar dependence structures, especially when $p=2$. To avoid problems with nearly non-identifiability in the bivariate case, 
we suggest that the two parameters, $\alpha_2^U$ and $\alpha_2^L$, be set to zero to prevent possible overparametrization. By doing this, we avoid possible convergence problems when estimating parameters and make the estimation faster as the likelihood function simplifies in this case. Simulation studies show that the reduced model with $\alpha_2^U = \alpha_2^L = 0$ fits data reasonably well even when the full model (\ref{mainmodel}) is used to simulate the data.

When $p > 2$, the composite likelihood approach can be used to estimate the parameters $\tht_{\Sigma}$ and $\tht_F$; see \cite{Varin.Vidoni2005} for an overview of composite likelihood methods. Let $f_{n,i_1,i_2}^{\Ww}$ be the pdf of $(W_{i_1,1},\ldots,W_{i_1,n},W_{i_2,1},\ldots,W_{i_2,n})^{\top}$ for $1 \leq i_1 < i_2 \leq p$. Then the composite pseudo-likelihood is
$$\ell^C(\mathbf{z}_1,\ldots,\mathbf
{z}_N) = \sum_{1 \leq i_1 < i_2 \leq p}\sum_{k=1}^N f_{n,i_1,i_2}(\mathbf{z}_{i_1,k},\mathbf{z}_{i_2,k};\tht_{F,i_1},\tht_{F,i_2},\tht_{\Sigma})
 -2 \sum_{i=1}^p\sum_{k=1}^N\sum_{j=1}^n  \log f_{1,i}^{\WW}(z_{ij,k};\tht_{F,i}).$$

Marginal densities $f_{n,i_1,i_2}$, $1 \leq i_1 < i_2 \leq p$, can be obtained in a simple form; see Appendix \ref{appx-pdf}. Parameter estimation can therefore be reasonably fast when $p > 2$.

%

\subsection{Goodness of fit of the estimated model and model misspecification}
\label{sec_MLE_gof}

To assess the goodness of fit of the estimated model, we 
compute the Spearman correlations, $S_{\rho}$, and the tail-weighted dependence measures of \cite{Krupskii.Joe2015}, $\varrho_L/\varrho_U$, for each pair of variables from the estimated model (to get model-based estimates). These quantities cannot be obtained in closed form for our model and we therefore simulate $50,000$ replicates from this model and then compute empirical estimates of these dependence measures from the simulated data set.

Simulation from the copula model (\ref{cop_pdf}) is straightforward. For given spatial locations, $s_1, \ldots, s_n$, the cross-covariance matrix $\Sigma_{\Zz}$ can be calculated using the selected cross-covariance function. One needs to generate a multivariate normal vector $\Zz$ with zero mean, unit variance and covariance matrix $\Sigma_{\Zz}$ and $2p+2$ i.i.d. $\mathrm{Exp}(1)$ exponential random variables $\ee_0^L, \ee_0^U, \ee_i^L, \ee_i^U$. Then (\ref{mainmodel}) can be used to compute $W_{ij}$ and finally obtain $U_{ij} = F_{1,i}^{W}(W_{i,j})$ for $i = 1,\ldots,p$ and $j=1,\ldots,n$. The joint dependence of $\Uu = (U_{11},\ldots,U_{1n},\ldots,U_{p1},\ldots,U_{pn})^{\top}$ is then given by the copula in (\ref{cop_pdf}).

The measures  $\varrho_L/\varrho_U$ can be used to estimate the strength of dependence in the lower/upper tails for a pair of variables. Unlike tail dependence coefficients which are defined as limiting quantities, accurate estimates of the tail-weighted measures can be obtained when the sample size is not large. We also compute these measures for the data set used to estimate the copula parameters (to get empirical estimates). We then compute the average (the absolute average) differences between the model-based and the corresponding empirical estimates of $S_{\rho}$, $\varrho_L$ and $\varrho_U$. We denote these averaged values by $\Delta_{\rho}, \Delta_L, \Delta_U$ ($|\Delta_{\rho}|, |\Delta_L|, |\Delta_U|$), respectively.

We now show that the proposed copula model with the exponential factors can provide an adequate fit to the data even when this model is misspecified. For illustration purposes, we consider the following two models with $p=2$:
\begin{eqnarray*}
\text{Model A: } \quad W_i &=& Z_i/\sqrt{V/\nu},\quad V\sim\chi^2(\nu),\\
\text{Model B: } \quad W_i &=& Z_i + \alpha_{i0}^U\pp_0^U + \alpha_{i}^U\pp_{i}^U - \alpha_{i0}^L\pp_0^L - \alpha_{i}^L\pp_i^L,
\end{eqnarray*}
where
\begin{equation} Z_i = \rho_i\tilde Z_0 + \sqrt{1-\rho_i^2}Z_i^*, \quad i=1,2, \label{LMC-Model-MLE} \end{equation}
and $Z_0, Z_1^*, Z_2^*$ are independent Gaussian processes with unit variance and the powered exponential covariance function, $C(d;\theta,\alpha)=\exp(-\theta d^{\alpha})$ ($\theta > 0, 0 < \alpha < 2$), with parameters $(\theta_0, \alpha_0), (\theta_1, \alpha_1)$ and $(\theta_2, \alpha_2)$, respectively. Here, $\chi^2(\nu)$ is the chi-squared distribution with $\nu > 0$ degrees of freedom and $\pp_0^U, \pp_0^L, \pp_1^U, \pp_1^L, \pp_2^U, \pp_2^L$ are independent Pareto random variables with the scale equal to 1 and shape equal to 4. It implies that Model B has very strong dependence in the tails; see \cite{Krupskii.Huser.ea2016}. We use $\tht = (\theta_0, \theta_1, \theta_2, \alpha_0, \alpha_1, \alpha_2, \rho_1, \rho_2)^{\top} = (0.25,0.35,0.45,0.3,0.4,0.3,0.6,0.8)^{\top}$ and $\nu = 4$ for Model A, corresponding to the Student-$t$ spatial process with moderate dependence in the tails, and $\tht = (0.55,0.65,0.75,1.1,1.2,1.3,0.6,0.8,)^{\top}$, $\alp = (\alpha_{10}^U, \alpha_{20}^U, \alpha_1^U, \alpha_2^U$, $\alpha_{10}^L, \alpha_{20}^L, \alpha_1^L, \alpha_2^L,)^{\top} = (1.1,1.3,0.5,0.0,0.8,0.9,0.6,0.0)^{\top}$ for Model B, corresponding to strong dependence, especially in the upper tail.

We now randomly select 10 data locations in $[0,1]^2$ and simulate two samples of size $N=1,000$, one from Model A and the other one from Model B. We then fit the misspecified copula model (\ref{mainmodel}) to the two simulated data sets and compute  $\Delta_{\rho}, \Delta_L, \Delta_U, |\Delta_{\rho}|, |\Delta_L|, |\Delta_U|$ for the estimated copula model for the two data sets to assess the adequacy of fit of this model; Table \ref{tab-gof-simdata} shows the results.

\begin{table}[h!]
\def~{\hphantom{0}}
\caption{{\footnotesize $\Delta_{\rho}, |\Delta_{\rho}|, \Delta_{L}, |\Delta_{L}|, \Delta_{U}, |\Delta_{U}|$ for models A and B. We simulated data from the estimated models A and B to calculate these values; we used $N = 1,000$ replicates. }}
\label{tab-gof-simdata}
\begin{center}
\begin{tabular}{lccc}
\hline
&$\Delta_{\rho}/ |\Delta_{\rho}|$ & $\Delta_{L}/ |\Delta_{L}|$ & $\Delta_{U}/ |\Delta_{U}|$  \\
\hline
&\multicolumn{3}{l}{Model A: Student-$t$ spatial process}\\
\hline
Variable 1 & ~0.00/0.02~ & ~0.07/0.07~ & $-$0.05/0.06~\\
Variable 2 & ~0.01/0.02~ & ~0.06/0.07~ & ~0.08/0.08~\\
Variables 1 and 2 & $-$0.01/0.02~ & ~0.06/0.08~ & ~0.00/0.05~\\
\hline
\hline
&\multicolumn{3}{l}{Model B: Pareto factors}\\
\hline
Variable 1 & $-$0.03/0.03~ & ~0.04/0.06~ & ~0.07/0.07~\\
Variable 2 & $-$0.04/0.04~ & ~0.08/0.08~ & ~0.05/0.05~\\
Variables 1 and 2 & $-$0.04/0.04~ & ~0.08/0.08~ & ~0.04/0.06~\\
\hline
\end{tabular}
\end{center}
\end{table}

We can see that the misspecified model (\ref{mainmodel}) with the exponential factors fits data generated from Model A and B quite well, both in the middle of the distribution and in its tails. Similar results hold for other sets of parameters and models for $(Z_1, Z_2)$. We obtained similar results with different locations in $[0,1]^2$ and with different sets of parameters for Models A and B. The copula model proposed in this paper can therefore be used for modeling multivariate spatial data with different dependence structures, even if the underlying spatial process is not described by model (\ref{base_model}) with the exponential factors.

\subsection{A conditional copula and interpolation}
\label{sec_MLE_sub3}

The estimated copula model is not restricted to a set of data locations but it also allows for interpolation of a spatial process at new sites where the values of variables are unknown. For a given covariance function of $\{Z_1(\ss),\ldots,Z_p(\ss)\}^{\top}$ and a new, $(n+1)$-th, location, one can obtain the new covariance matrix $\Sigma_{Z}$ of the measurements of this Gaussian process at known sites and the new location. This covariance matrix can be used to compute the joint density $c_{n+1,p}^{\Ww}$ and then to construct the conditional distribution used for interpolation on the uniform scale. Estimated marginal distributions can then be used to transform the interpolated values to the original scale as we show below.

Let $\widehat \tht_F$, $\widehat \tht_{\Sigma}$ be estimates of $\tht_F$ and $\tht_{\Sigma}$, respectively. For a given vector of data $(\mathbf{u}_1,\ldots,\mathbf{u}_p)^{\top}$, with $\mathbf{u}_i = (u_{i,1},\ldots,u_{i,n})^{\top}$ for $i=1,\ldots,p$, and the new vector (corresponding to a new, $(n+1)$-th, location), $\mathbf{u}_0 =(u_{1,n+1},\ldots,u_{p,n+1})^{\top}$, on the uniform scale, we can obtain the following conditional distribution:
$$\widehat C_{0|n,p}^{\Ww}(\mathbf{u}_0|\mathbf{u}_1,\ldots,\mathbf{u}_p) := \frac{\int_0^{\mathbf{u}_0} c_{n+1,p}^{\Ww}(\mathbf{u}_1,u_{1,n+1}^*,\ldots,\mathbf{u}_p,u_{p,n+1}^*;\widehat \tht_F, \widehat \tht_{\Sigma}) \d \mathbf{u}_0^*}{c_{n,p}^{\Ww}(\mathbf{u}_1,\ldots,\mathbf{u}_p;\widehat \tht_F, \widehat \tht_{\Sigma})},$$
where $\mathbf{u}_0^* = (u_{1,n+1}^*,\ldots,u_{p,n+1}^*)^{\top}$.  
The conditional distribution for the $i$-th variable is
$$
\widehat C_{i,0|n,p}^{\Ww}(u_{i,n+1}|\mathbf{u}_1,\ldots,\mathbf{u}_p) = \widehat C_{0|n,p}^{\Ww}(\mathbf{u}_0|\mathbf{u}_1,\ldots,\mathbf{u}_p), \quad \text{ with } u_{k, n+1} = 1 \text{ for } k \neq i.
$$
Using this conditional distribution, we can calculate different quantities of interest, including the conditional expectation, $\widehat m_i$, or the conditional median, $\widehat q_{0.5,i}$, of the $i$-th variable:
$$
\widehat m_i := \int_0^1 u_{i,n+1}^* \,\widehat c_{i,0|n,p}^{\Ww}(u_{i,n+1}^*,\mathbf{u}_1,\ldots,\mathbf{u}_p)\,\d u_{i,n+1}^*, \quad \widehat q_{0.5,i} := (\widehat C_{i,0|n,p}^{\Ww})^{-1}(0.5|\mathbf{u}_1,\ldots,\mathbf{u}_p),
$$
where
$$
\widehat c_{i,0|n,p}^{\Ww}(u_{i,n+1}^*,\mathbf{u}_1,\ldots,\mathbf{u}_p) = \frac{\p \widehat C_{i,0|n,p}^{\Ww}(u_{i,n+1}| \mathbf{u}_1,\ldots,\mathbf{u}_p)}{\p u_{i,n+1}}\Bigg|_{u_{i,n+1} = u_{i,n+1}^*}\,.
$$
Here, numerical integration can be used to compute $C_{0|n,p}^{\Ww}(\mathbf{u}_0|\mathbf{u}_1,\ldots,\mathbf{u}_p)$. 

In applications, however, a spatial process usually needs to be interpolated on the original scale. For that reason, one needs to estimate the marginal distributions of the process at each location and to use these distributions to convert the interpolated values to the original scale. If $\widehat G_{i,s}$ is the estimated univariate marginal distribution function for the $i$-th variable at location $s$, then we can transform the uniform data to the original scale. For example, the predicted median at location $s$ on the original scale will be $\widehat z_{0.5;i,s} = \widehat G_{i,s}^{-1}(\widehat q_{0.5,i})$.

\section{Application to temperature and atmospheric pressure data}
\label{sec_empstudy}

\subsection{Data and marginal models}
\label{sec_empstudy_data}

In this section, we apply our model to estimate the joint dependence structure of daily mean temperature (\texttt{TAVG}, average of all 5-minute averaged temperature observations each day) and daily mean atmospheric pressure (\texttt{PAVG}, average of all 5-minute averaged station air pressure observations each day) readings in Oklahoma, USA. The data are available at \texttt{mesonet.org}. The aim of this study is to estimate the proposed copula model using the observed values at several stations and then interpolate data at new locations. We consider 17 weather stations in the central part of the state: Acme, Apache, Chandler, Chickasha, Fort Cobb, Guthrie, Hinton, Marena, Minco, Ninnekah, OKC East, OKC North, Perkins, Shawnee, Stillwater, Washington and Watonga. These stations are located close to each other with the maximum distance between two stations being 170 kilometers. The area of interest has no big mountains and the weather conditions remain consistent across the area. The proposed copula model (\ref{base_model}) may therefore be suitable for modeling the joint dependence of temperature and pressure data when unobserved factors (common weather patterns) affect the temperature and pressure at all stations in this area. 

Weather patterns can change in winter and therefore we selected observations from May 1st to September 30th, 2015, 153 days in total. We use 13 stations (Acme, Apache, Chandler, Chickasha, Fort Cobb, Guthrie, Hinton, Ninnekah, OKC East, Perkins, Shawnee, Stillwater and Washington) to fit the model and 4 stations (Marena, Minco, OKC North and Watonga) to interpolate data for all 153 days and compare the interpolated values against the observed ones.

To remove serial dependence, we fitted the autoregressive model with 6 lags for both temperature and pressure data. We also included a quadratic trend in the model for the temperature data, as temperatures are usually higher in July and August. We found that including longitude as a spatial covariate improved the fit of the marginal model for the pressure data but not for the temperature data. One possible reason for spatial covariates to have no significant effect on the temperature data is the proximity of the weather stations and a larger variability of these data depending on the orography. We therefore can write the models for univariate marginals as follows:
\begin{equation}
\label{AR-model}
\begin{aligned}
\emph{temp}_{\,s,t} &= \beta_0^1 + \beta_1^1 t + \beta_2^1 t^2 + \sum_{m=1}^{6}\alpha_{m}^1\,\emph{temp}_{\,s,t-m}\, + \epsilon_{s,t}\,, \\
\emph{prss}_{\,s,t} &= \beta_0^2 + \beta_1^2\,\emph{lon}_s + \sum_{m=1}^{6}\alpha_{m}^2\,\emph{prss}_{\,s,t-m}\, + \eta_{s,t}\,,
\end{aligned}
\end{equation}
where $\emph{temp}_{\,s,t}$ and $\emph{prss}_{\,s,t}$ are average temperature and pressure, respectively, measured at station $s$ at day $t$; $\emph{lon}_s$ is  longitude of station $s$, $s=1,\ldots,13$ and $t=1,\ldots,153$. We assume that $\epsilon_{s,t}$ and $\eta_{s,t}$ are i.i.d. random variables for each location $s$ and for each day $t$ and we found that the skew-$t$ distribution of \cite{Azzalini.Capitanio2003} and the normal distribution, respectively, fit residuals $\epsilon_{s,t}$ and $\eta_{s,t}$ quite well. We checked the fitted residuals for uncorrelatedness using the Ljung-Box test.

\subsection{Preliminary diagnostics of the data set}
\label{sec_empstudy_diagn}

\begin{figure}
\begin{center}
\includegraphics[width=7in,height=7in]{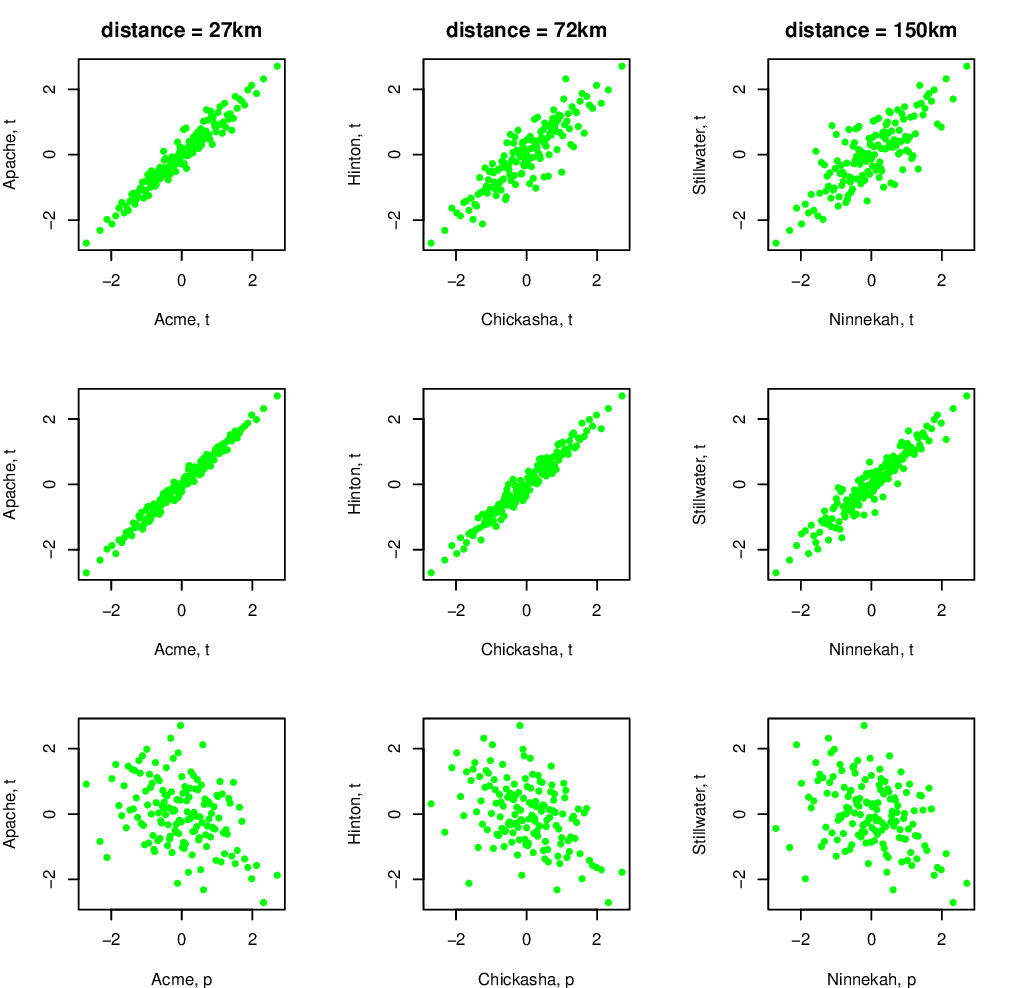}
\caption{{\footnotesize Scatter plots of normal scores for temperature data (top), pressure data (middle), pressure (x-axis) and temperature (y-axis) data (bottom) for Acme, Apache (left), Chickasha, Hinton (middle), and Ninnekah, Stillwater (right) stations.}}
\label{fig_mvn2_stations}
\end{center}
\end{figure}

We convert the fitted residuals, $\widehat\epsilon_{s,t}$ and $\widehat\eta_{s,t}$, from the marginal models to uniform scores. For $s=1,\ldots,13$, we define
$$
u^1_{s,t} = \{\texttt{rank}(\widehat\epsilon_{s,t}) - 0.5\}/153, \quad u^2_{s,t} = \{\texttt{rank}(\widehat\eta_{s,t}) - 0.5\}/153, \quad t=1,\ldots,153.
$$
If the marginal models fit the data well, the uniform scores, $u^1_{s,t}, u^2_{s,t}$, $t=1,\ldots,153$, should have an approximate $U(0,1)$ distribution for any $s=1,\ldots,13$. We can therefore convert them to the normal scores using the inverse standard normal distribution function:
$$
z^1_{s,t} = \Phi^{-1}(u^1_{s,t}), \quad  z^2_{s,t} = \Phi^{-1}(u^2_{s,t}), \quad t=1,\ldots,153.
$$
Under the assumption of joint normality, the vector $\mathbf{z}_t = (z_{1,t}^1,\ldots,z_{17,t}^1, z_{1,t}^2,\ldots,z_{17,t}^2)^{\top}$ has a multivariate normal distribution. We can therefore draw the scatter plots for each pair of variables from the vector $\mathbf{z}_t$ (corresponding to repeated daily mean temperature or pressure measurements at two spatial locations) to check if these plots have the expected elliptical shape for a bivariate normal distribution. The use of the normal scores scatter plots to detect departures from normality has been advocated by \cite{Nikoloulopoulos.Joe.ea2012}. We draw the normal scores scatter plots for some pairs of the normal scores in Fig. \ref{fig_mvn2_stations}.

Sharp tails in the normal scores scatter plots of the temperature data indicate that the dependence in the tails of these data is stronger than that of the normal distribution. The sharper lower tails of these scatter plots also indicate that the dependence in the temperature data might be stronger in the lower tail. The figure shows the negative dependence between the normal scores of the temperature and pressure, with a sharper lower right tail, so that the joint distribution of $z^1_{s,t}$ and $z^2_{s,t}$ is clearly asymmetric. Models based on multivariate normality are therefore not suitable for modeling the joint dependence of the temperature and pressure data.

To confirm these findings, we compute the Spearman correlations and the tail-weighted dependence measures, $\varrho_L/\varrho_U$, 
for each pair of variables from the vector $\mathbf{u}_t = (u_{1,t}^1,\ldots,u_{17,t}^2,1-u_{1,t}^2,\ldots,1-u_{17,t}^2)^{\top}$, as explained in Section \ref{sec_MLE_gof}. We use the reflected pressure data, $1-u_{s,t}^2$, to get positive dependence between the temperature and pressure data. However, the proposed model can also handle negative dependencies if the tail coefficients $\alpha_{i0}^L, \alpha_{i0}^U, \alpha_{i}^L, \alpha_{i}^U$ in (\ref{mainmodel}) are negative for some variables $i$. In particular,  for the temperature and pressure data set, it is equivalent to using the original (not reflected data) and negative coefficients $\alpha_{20}^L, \alpha_{20}^U, \alpha_{2}^L, \alpha_{2}^U$.

Let $\varrho_N(u_1,u_2)$ be the value $\varrho_L(u_1,u_2) = \varrho_U(u_1,u_2)$ for data generated from the bivariate normal copula with the Spearman's $\rho$ equal to $\Cor(u_1,u_2)$. If the bivariate copula for the pair $(u_1,u_2)$ is a normal copula, we expect to get close values for $\varrho_L, \varrho_U$ and $\varrho_N$. If the dependence in the lower (upper) tail is stronger than that for the normal copula, then we expect the value of $\varrho_L$ ($\varrho_U$) to be larger than $\varrho_N$. For $i=1,2$, we compute:
\begin{eqnarray*}
S_{\rho}^i &=& \sum_{s_1<s_2}\Cor(u_{s_1,t}^i,u_{s_2,t}^i)/136, \quad S_{\rho}^{12} = \sum_{s_1\leq s_2}\Cor(u_{s_1,t}^1,1-u_{s_2,t}^2)/153,\\
\varrho_N^i &=& \sum_{s_1<s_2}\varrho_N(u_{s_1,t}^i,u_{s_2,t}^i)/136, \quad \ \varrho_N^{12} = \sum_{s_1\leq s_2}\varrho_N(u_{s_1,t}^1,1-u_{s_2,t}^2)/153, \\
\varrho_L^i &=& \sum_{s_1<s_2}\varrho_L(u_{s_1,t}^i,u_{s_2,t}^i)/136, \quad \ \varrho_L^{12} = \sum_{s_1\leq s_2}\varrho_L(u_{s_1,t}^1,1-u_{s_2,t}^2)/153, \\
\varrho_U^i &=& \sum_{s_1<s_2}\varrho_U(u_{s_1,t}^i,u_{s_2,t}^i)/136, \quad \ \varrho_U^{12} = \sum_{s_1\leq s_2}\varrho_U(u_{s_1,t}^1,1-u_{s_2,t}^2)/153. \\
\end{eqnarray*}
\vspace{-4em}

Here, the superscripts $1, 2, 12$ indicate that the calculated measures are averaged for all pairs of different locations for variable 1 (temperature), variable 2 (pressure) and variables 1 and 2 (to measure cross dependencies). The results are presented in Table \ref{tab-diag-mvn2}.

\begin{table}
\def~{\hphantom{0}}
\caption{{\footnotesize $S_{\rho}^i, \varrho_N^i, \varrho_L^i, \varrho_U^i$ for $i=1,2$ and $i=1$ and 2
(standard errors are shown in parentheses)}}
\label{tab-diag-mvn2}
\begin{center}
\begin{tabular}{lcccc}
\hline
Variable &$S_{\rho}$ & $\varrho_N$ & $\varrho_L$ & $\varrho_U$  \\
\hline
Variable 1 & 0.85 & 0.71 & 0.85(0.06) & 0.79(0.06)\\
Variable 2 & 0.97 & 0.94 & 0.95(0.03) & 0.92(0.02)\\
Variables 1 and 2 & 0.36 & 0.17 & 0.59(0.16) & 0.10(0.13)\\
\hline
\end{tabular}
\end{center}
\end{table}
We see that the dependence for variable 1 is stronger than it is for the normal copula in the lower tail. In addition, the cross dependence between variable 1 and (reflected) variable 2 is much stronger in the upper tail. We therefore need a model that can handle tail dependence and asymmetric dependence for the temperature and pressure data.

\subsection{Estimating the joint dependence}
\label{sec_empstudy_MLE}

We assume that the residuals $\emph{temp}_{\,s,t_1}$ and $\emph{prss}_{\,s,t_2}$ are independent for $t_1 \neq t_2$. This is a plausible assumption as the autocorrelation plots show no significant correlations at lags from 1 to 20. The vectors of residuals, $(\emph{temp}_{\,s,t},\emph{prss}_{\,s,t})$, can therefore be treated as replicates for $t=1,\ldots,153$. We apply the model (\ref{mainmodel}) to the residuals obtained from the marginal models in Section \ref{sec_empstudy_data} transformed to uniform scores $u_{s,t}^1$ and $1-u_{s,t}^2$, $s=1,\ldots,17$ and $t=1,\ldots,153$. Before fitting the model, we need to model the cross-covariance of $\Zz = (Z_{1,1},\ldots,Z_{1,17},Z_{2,1},\ldots,Z_{2,17})^{\top}$. We select the same linear model of coregionalization (\ref{LMC-Model-MLE}) 
as in Section \ref{sec_MLE_gof}. 
Different models can be used to model the covariance structure of $Z$, including the bivariate Mat\'{e}rn model; we found however that these models did not significantly change the results reported in this section.
%

We set two parameters, $\alpha_2^L, \alpha_2^U$, to zero as discussed in Section \ref{sec_MLE_sub3} to avoid convergence problems with the algorithm and to increase the speed of computation. With this restriction, the maximum likelihood estimates were obtained in about five minutes on a Core i5-2410M CPU@2.3 GHz.
To assess the goodness of fit of the estimated copula model (\ref{mainmodel}) with the exponential factors (Model 1), we computed $\Delta_{\rho}, \Delta_L, \Delta_U, |\Delta_{\rho}|, |\Delta_L|, |\Delta_U|$ as explained in Section \ref{sec_MLE_gof}. If Model 1 fits the data well, we expect these values to be small. 
For comparison, we also fit the model (\ref{LMC-Model-MLE}) without exponential factors (Model 2, assuming $\alpha_{10}^U = \alpha_1^U = \alpha_{10}^L = \alpha_1^L = \alpha_{20}^U = \alpha_2^U = \alpha_{20}^L = \alpha_2^L = 0$ in (\ref{mainmodel})). Model 2 assumes the normal copula with no tail dependence. The results are presented in Table \ref{tab-gof-data}.

\begin{table}[h!]
\def~{\hphantom{0}}
\caption{{\footnotesize $\Delta_{\rho}, |\Delta_{\rho}|, \Delta_{L}, |\Delta_{L}|, \Delta_{U}, |\Delta_{U}|$ for models 1 and 2. We simulated data from the estimated models 1 and 2 to calculate these values; we used $N = 100,000$ replicates. }}
\label{tab-gof-data}
\begin{center}
\begin{tabular}{lccc}
\hline
&$\Delta_{\rho}/ |\Delta_{\rho}|$ & $\Delta_{L}/ |\Delta_{L}|$ & $\Delta_{U}/ |\Delta_{U}|$  \\
\hline
&\multicolumn{3}{c}{Model 1; BIC = $-$11796}\\
\hline
Variable 1 & $-$0.03/0.03~ & $-$0.04/0.04~ & $-$0.02/0.08~\\
Variable 2 & ~\,0.00/0.01~ & $-$0.01/0.01~ & ~\,0.00/0.01~\\
Variables 1 and 2 & $-$0.10/0.10~ & ~\,0.12/0.12~ & ~\,0.07/0.08~\\
\hline
\hline
&\multicolumn{3}{c}{Model 2; BIC = $-$11565}\\
\hline
Variable 1 & $-$0.03/0.03~ & ~0.10/0.10~ & ~\,0.03/0.06~\\
Variable 2 & ~\,0.00/0.01~ & ~0.02/0.02~ & $-$0.01/0.02~\\
Variables 1 and 2 & $-$0.02/0.03~ & ~0.41/0.41~ & $-$0.10/0.12~\\
\hline
\end{tabular}
\end{center}
\end{table}
We can see that both Model 1 and Model 2 fit the covariance structure quite well; however, Model 2 (with no exponential factors) significantly underestimates the cross dependence in the lower tail. Model 1 significantly improves the fit in the lower tail. 
We obtained very similar results with different choices of the cross-covariance function used to construct Models 1 and 2.

\subsection{Interpolating data at locations with unknown values of variables}
\label{sec_empstudy_interpol}

We now interpolate data for stations Marena, Minco, OKC North and Watonga. For a given location $s$, we use the average values observed at the three nearest stations as starting points for $temp_{s,t}$ and $prss_{s,t}$ for $t=1,\ldots,6$. Assume we have obtained interpolated medians of temperature and pressure values at time $t \geq 6$. We compute the interpolated medians of these two variables on the uniform $(0,1)$ scale at location $s$ and time $t+1$ using the vector of uniform scores $\mathbf{u}_t$ as explained in Section \ref{sec_MLE_sub3}. We convert the interpolated medians to residuals, $\epsilon_{s,t+1}$ and $\eta_{s,t+1}$, using the inverse skew-$t$ and normal distributions, respectively, and then compute the interpolated medians at time $t+1$, $temp_{s,t+1}$ and $prss_{s,t+1}$, using (\ref{AR-model}). We repeat this procedure for all $t \leq 153$.

\begin{figure}[t!]
\begin{center}
\includegraphics[width=6in,height=3in]{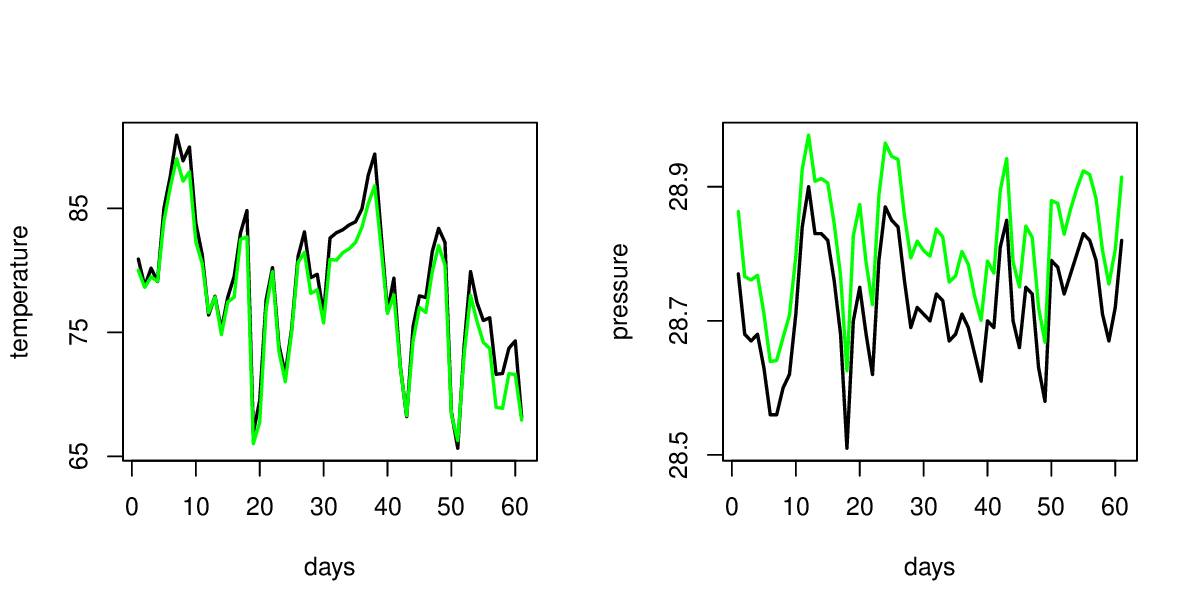}
\caption{{\footnotesize Interpolated mean daily temperature (degrees Fahrenheit) and pressure values (inches of mercury) for station OKC North (green line) and observed values (black line), from August, 1 to September, 30.}}
\label{interpol1}
\end{center}
\end{figure}

We also interpolate the 5\% and 95\% quantiles of the temperature and pressure variables. To obtain interpolated quantiles at time $t$, we compute the interpolated medians of these variables at time $t-1$ and the interpolated 5\% and 95\% quantiles of these two variables on the uniform $(0,1)$ scale at time $t$ using the vector of uniform scores $\mathbf{u}_t$.  We then apply (\ref{AR-model}) to obtain the 5\% and 95\% quantiles of these variables on the original scale at time $t$. Fig. \ref{interpol1} shows the interpolated and observed medians for temperature and pressure for station OKC North for the last two months, August and September, 2015.


We see that variability in time is captured quite well by the model. We next compute the mean absolute errors for the four stations for interpolated values obtained using Model 1. For comparison, we do the same for the temperature and pressure values obtained by taking the average of values measured at the three nearest stations. Table \ref{tab-interpol-MAE} shows the results. For the pressure data, both methods give comparable results. For the temperature data, the average of the three nearest neighbors gives slightly better results if these neighbors are close to the new location (Marena and OKC North stations). If the new location is far from the stations with the observed data, Model 1 gives better results.

\begin{table}
\caption{{\footnotesize Mean absolute errors for 4 stations: Marena, OKC North and Minco, Watonga. The interpolated values are calculated using Model 1/average of three nearest neighbors.}}
\label{tab-interpol-MAE}
\begin{tabular}{lccccc}
\hline
Station & Marena & Minco & OKC North & Watonga \\
\hline
Pressure & 0.08/0.10 & 0.17/0.20 & 0.09/0.04 & 0.27/0.23\\
Temperature & 0.65/0.61 & 1.24/1.42 & 1.20/1.11 & 0.92/1.20\\
\hline
\end{tabular}
\end{table}

\begin{figure}[t!]
\begin{center}
\includegraphics[width=6.3in,height=3.4in]{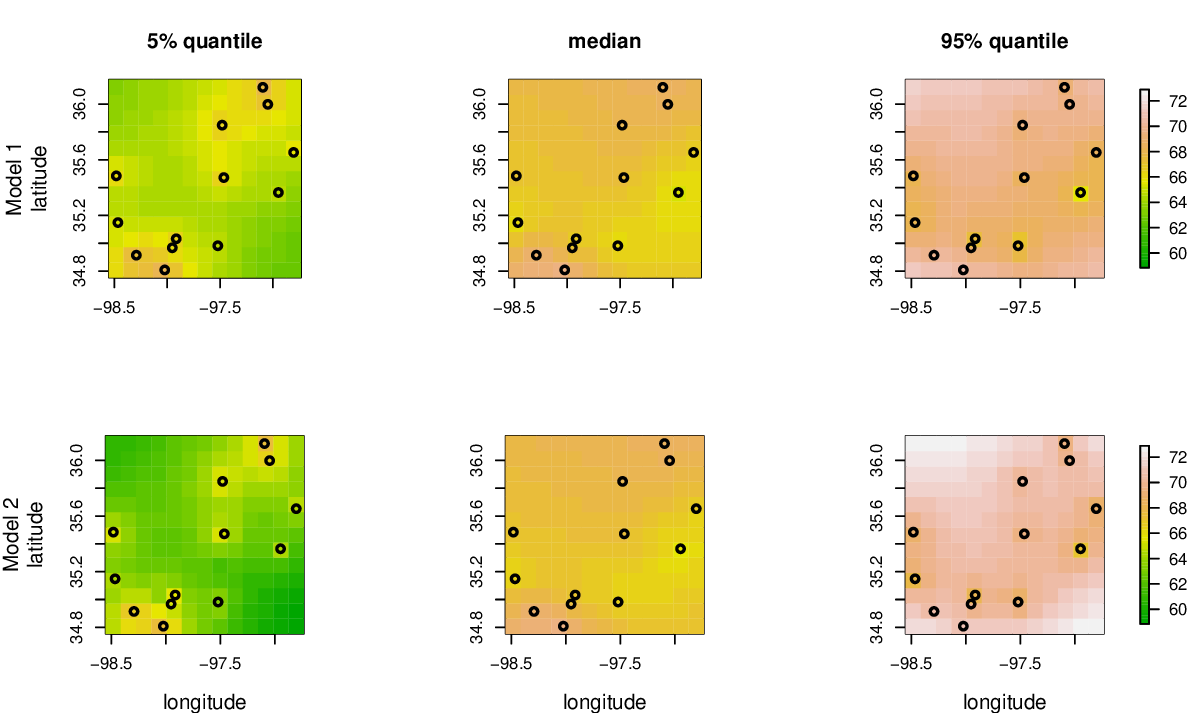}
\caption{{\footnotesize The predicted 5\% (left), 50\% (middle) and 95\% (right) quantiles for Model 1 (top) and Model~2 (bottom) for mean daily temperatures in the area of study (degrees Fahrenheit), calculated for September 19, 2015. The 13 stations with recorded temperature data are shown as circles.}}
\label{heatmap1}
\end{center}
\end{figure}

Finally, for the area of study, between $34.8^{\circ}$ and $36.2^{\circ}$ North and between $96.7^{\circ}$ and $98.5^{\circ}$ West, we compute the predicted medians as well as the 5\% and 95\% quantiles for the temperature data for September 19, 2015. The observed temperatures at this day were very low and the pressure readings were very high. The interpolated values both for Model 1 and 2 are shown in Fig. \ref{heatmap1}. The predicted medians are similar for the two models, however the 5\% quantiles are higher and the 95\% quantiles are lower for Model 1. This model can handle strong dependence between low temperature and high pressure values and the predicted uncertainty for the new locations is smaller when taking the pressure readings into account. Model 2 assumes nearly independence of very low temperature and very high pressure values (this model underestimates the joint dependence between low temperatures and high pressure values as follows from Table \ref{tab-gof-data}) and therefore the pressure data do not help to reduce the predicted uncertainty in this case.

\section{Conclusion}
\label{sec_discussion}

We proposed a new copula model for multivariate spatial data that can handle tail dependence and asymmetric dependence within each variable as well as between different variables. This model is a generalization of any model based on multivariate normality. The widely used linear model of coregionalization is an example. Parameters in the model can be estimated by likelihood and the formula for the joint copula density is quite simple so that parameters can be computed fairly easily. The model allows simple interpretation when factors affecting the joint dependence of multivariate spatial data exist. 

In this paper, we assumed that data for all variables are known at the data locations. However, data for some variables can be missing at some locations. To estimate the copula parameters in this case, one needs to use some imputation methods or the modified likelihood function that takes into account all the available information. This is a topic of future research.

While the proposed model can generate a wide range of dependence structures, it assumes a linear structure when some exponential factors are added to the  multivariate Gaussian process. One direction for future research can therefore be to consider models with multiplicative factors or more general models based on the multivariate process:
\begin{equation}
W_i(\ss) = f_i\{Z_i(\ss), \ee_0, \ee_i\}, \quad i = 1,\ldots,p, \label{extnd_model}
\end{equation}
where $\{Z_1(\ss),\ldots,Z_p(\ss)\}^{\top}$ is a multivariate Gaussian process and $\ee_0, \ee_1, \ldots, \ee_p$ are factors that do not depend on the location $\ss$ (not necessarily exponential), introduced to increase the flexibility of the model. The choice of functions $f_1,\ldots,f_p$ and distributions of factors $\ee_0, \ee_1,\ldots,\ee_p$ can define the dependence properties of the joint distribution and the resulting copula. Another direction for future research is to study anisotropic factor copula models when the joint dependence of the variables is affected by a location or some other factors. One way to introduce anisotropy can be, for example, by spatially varying coefficients $\alpha_{i0}^L, \alpha_{i}^L, \alpha_{i0}^U$ and $\alpha_{i}^U$, $i=1,\ldots,p$, in (\ref{base_model}) or functions $f_1,\ldots,f_p$ in (\ref{extnd_model}).

\bibliographystyle{apalike}
\bibliography{fmcop}

\setcounter{section}{1}
\renewcommand{\thesection}{\Alph{section}}
\section*{Appendix}

\subsection{Formula for $f_{n,p}^{\Ww}$ for $p=2$} \label{appx-pdf}
Let $\mathbf{w} = (\mathbf{w}_1, \mathbf{w}_2)^{\top}$, where $\mathbf{w}_i = (w_{i1},\ldots,w_{in})^{\top}$, $i=1,2$, and let $\Sigma_{\Zz}$ be a covariance matrix of the vector $\Zz = (Z_{11},\ldots,Z_{1n}, Z_{21},\ldots,Z_{2n})^{\top}$ as defined in (\ref{mainmodel}). One can show that the joint density for vector $\mathbf{W}^* = (W_{11}^*,\ldots,W_{1n}^*, W_{21}^*,\ldots,W_{2n}^*)^{\top}$, where $W_{ij}^* = Z_{ij} + \alpha_{i0}^U\ee_0^U - \alpha_{i0}^L\ee_0^L$, is
$$
f_{n,2}^{\Ww^*}(\mathbf{w}_1,\mathbf{w}_2) =  K_{\mathbf{w}}\exp\left(\frac{c_1^2c_{22}+2c_1c_2c_{12}+c_2^2c_{11}}{2c_{\delta}}\right)\Phi_{\rho^*}\left\{\frac{c_1c_{22}+c_2c_{12}}{(c_{\delta}c_{22})^{1/2}}, \frac{c_1c_{12}+c_2c_{11}}{(c_{\delta}c_{11})^{1/2}}\right\},
$$
where $c_{\delta} = c_{11}c_{22} - c_{12}^2$, $K_{\mathbf{w}}=(2\pi)^{1-n}\{c_{\delta}\det(\Sigma)\}^{-1/2}\exp(-\mathbf{w}^{\top}\Sigma^{-1}\mathbf{w}/2)$, $\Phi_{\rho^*}$ is the cumulative distribution function of a bivariate standard normal random variable with correlation $\rho^*$ and
$$
c_1 = c_1(\mathbf{w}) = \alpha_{10}^Us_1(\mathbf{w})+\alpha_1^Us_2(\mathbf{w})-1, \quad c_2 = c_2(\mathbf{w}) = -\alpha_{10}^Ls_1(\mathbf{w})-\alpha_1^Ls_2(\mathbf{w})-1,
$$
$$
c_{11}=(\alpha_{10}^U)^2s_{11}+2\alpha_{10}^U\alpha_1^Us_{12}+(\alpha_1^U)^2s_{22},\quad
c_{22}=(\alpha_{10}^L)^2s_{11}+2\alpha_{10}^L\alpha_1^Ls_{12}+(\alpha_1^L)^2s_{22},
$$
$$
c_{12}=\alpha_{10}^U\alpha_{10}^Ls_{11}+(\alpha_{10}^U\alpha_1^L+\alpha_{10}^L\alpha_1^U)s_{12}+\alpha_1^U\alpha_1^Ls_{22},
$$
$$
s_1 = s_1(\mathbf{w}) = \sum_{j=1}^n(\Sigma_{\Zz}^{-1}\mathbf{w})_j, \quad s_2 = s_2(\mathbf{w}) = \sum_{j=n+1}^{2n}(\Sigma_{\Zz}^{-1}\mathbf{w})_j,
$$
$$
s_{11} = \sum_{j_1,j_2=1}^n(\Sigma_{\Zz}^{-1})_{j_1,j_2}, \quad s_{22} = \sum_{j_1,j_2=n+1}^{2n}(\Sigma_{\Zz}^{-1})_{j_1,j_2}, \quad s_{12} = \sum_{j_1=1}^n\sum_{j_2=n+1}^{2n}(\Sigma_{\Zz}^{-1})_{j_1,j_2}.
$$

The density of $V_i = \alpha_{i}^U\ee_0^U - \alpha_{i}^L\ee_0^L$ is $f_{V_i}(w) = [\exp\{-w_+/\alpha_i^U-(-w)_+/\alpha_i^L\}]/(\alpha_i^U+\alpha_i^L)$. We use the convolution formula to get
$$
f_{n,2}^{\Ww}(\mathbf{w}_1,\mathbf{w}_2) = \int_{\mathbb{R}^2}f_{n,2}^{\Ww^*}(\mathbf{w}_1-v_1,\mathbf{w}_2-v_2)f_{V_1}(v_1)f_{V_2}(v_2)\d v_1 \d v_2.
$$
If we assume that $\alpha_2^U = \alpha_2^L = 0$, the formula simplifies to a one-dimensional integral:
$$
f_{n,2}^{\Ww}(\mathbf{w}_1,\mathbf{w}_2) = \int_{\mathbb{R}^1}f_{n,2}^{\Ww^*}(\mathbf{w}_1-v_1,\mathbf{w}_2)f_{V_1}(v_1)\d v_1
$$
$$
= \frac{1}{\alpha_1^U+\alpha_1^L}\,\int_{\mathbb{R}^1_+}\left\{f_{n,2}^{\Ww^*}(\mathbf{w}_1-v_1,\mathbf{w}_2)\exp(-v_1/\alpha_1^U) + f_{n,2}^{\Ww^*}(\mathbf{w}_1+v_1,\mathbf{w}_2)\exp(-v_1/\alpha_1^L) \right\} \d v_1.
$$
This integral can be evaluated with very good accuracy via Gauss-Legendre quadrature using $25-30$ quadrature points; see \cite{Stroud.Secrest1966} for details.

\subsection{Proof of Proposition \ref{prop-1}} \label{appx-prop1}

For simplicity, we omit indices for the correlation coefficient $\rho=\rho_{1,2}^{1:2}$. We have:
$$
F_{2,1:2}^{\Ww}(z_1,z_2) = \int_{\mathbb{R}^3_+}\Phi_{\rho}(z_1-\alpha_{10}^Uv_0-\alpha_1^Uv_1, z_2-\alpha_{20}^Uv_0-\alpha_2^Uv_2)\exp(-v_0-v_1-v_2)\d v_0
\d v_1\d v_2\,.
$$
We use the integration by parts formula with respect to $v_1$ and $v_2$ to find that
\begin{equation} \label{appx2-base}
F_{2,1:2}^{\Ww}(z_1,z_2) = \int_{\mathbb{R}^1_+}\{I_0(v_0)-I_1(v_0)-I_2(v_0)+I_{12}(v_0)\}\d v_0,
\end{equation}
where $I_0(v_0) = \Phi_{\rho}(z_1-\alpha_{10}^Uv_0,z_2-\alpha_{20}^Uv_0)\exp(-v_0)$, $I_1(v_0)=\Phi_{\rho}(z_1-\alpha_{10}^Uv_0-\rho/\alpha_{2}^U,z_2-\alpha_{20}^Uv_0-1/\alpha_{2}^U)\exp\{(\delta_2-1)v_0+0.5/(\alpha_{2}^U)^2-z_2/\alpha_{2}^U\}$,
$I_2(v_0)=\Phi_{\rho}(z_1-\alpha_{10}^Uv_0-1/\alpha_{1}^U,z_2-\alpha_{20}^Uv_0-\rho/\alpha_{1}^U)\exp\{(\delta_1-1)v_0+0.5/(\alpha_{1}^U)^2-z_1/\alpha_{1}^U\}$,
$I_{12}(v_0)=\Phi_{\rho}(z_1-\alpha_{10}^Uv_0-1/\alpha_{1}^U-\rho/\alpha_{2}^U,z_2-\rho/\alpha_{2}^U-\alpha_{20}^Uv_0-1/\alpha_{2}^U)\exp\{(\delta_{12}-1)v_0+0.5(\rho_{12}^*)^2-z_1/\alpha_{1}^U-z_2/\alpha_{2}^U\}$,
and $(\rho_{12}^*)^2 = \{(\alpha_1^U)^2+2\rho\alpha_1^U\alpha_2^U+(\alpha_2^U)^2\}/(\alpha_1^U\alpha_2^U)^2$.

For the marginal distribution, $F_{1,i}^{W}(z) = \Phi(z) - [\alpha_{i0}^U\exp\{-z/\alpha_{i0}^U+0.5/(\alpha_{i0}^U)^2\}\Phi(z-1/\alpha_{i0}^U) - \alpha_{i}^U\exp\{-z/\alpha_{i}^U+0.5/(\alpha_{i}^U)^2\}\Phi(z-1/\alpha_{i}^U)]/(\alpha_{i0}^U-\alpha_i^U)$, $i=1,2$. Let $\delta_i \neq 1$, $\alpha_i^* = \max(\alpha_{i0}^U, \alpha_i^U)$ and let $z_i=c_i+\alpha_i^*\log n$, where $c_i = 0.5/\alpha_i^*+\alpha_i^*\log(\alpha_i^*/|\alpha_{i0}^U-\alpha_i^U|) - \alpha_i^*\log x_i$. This implies that $F_{1,i}^{W}(z_i) = 1 - x_i/n + o(1/n)$.

\emph{Case 1:} $\delta_1 > 1, \delta_2 > 1$. We get
\begin{multline}\int_{\mathbb{R}^1_+}I_0(v_0)\d v_0 = \int_{\mathbb{R}^1_+}\Phi_{\rho}\{c_1-\alpha_{10}^U(v_0-\log n),c_2-\alpha_{20}^U(v_0-\log n)\}\exp(-v_0)\d v_0 \\
= 1 - \frac{1}{n}\sum_{k=1}^2\alpha_{k0}^U\int_{\mathbb{R}^1}\Phi\left\{\frac{c_{3-k}-\rho c_k + (\alpha_{3-k,0}^U-\rho\alpha_{k0}^U)v}{(1-\rho^2)^{1/2}}\right\}\phi(c_k+\alpha_{k0}^Uv)\exp(v)\d v + o\left(\frac{1}{n}\right). \label{appx2-I0}
\end{multline}
To compute the integrals in (\ref{appx2-I0}), we use the following equality:
\begin{equation} \label{appx2-intf}
\int_{\mathbb{R}^1}\exp(\theta v)\phi(v)\Phi(q v)\d v = \exp(0.5\theta^2)\Phi\left\{\frac{\theta q}{(q^2+1)^{1/2}}\right\}.
\end{equation}
This equality can be obtained by differentiating the integral on the left-hand side with respect to the parameter $q$. We apply (\ref{appx2-intf}) to (\ref{appx2-I0}) and, after combining all terms, we get:
$$
\int_{\mathbb{R}^1_+}I_0(v_0)\d v_0 = 1 - \frac{y_1}{n}\Phi\left\{\frac{\rho_{12}}{2}+\frac{\log(y_1/y_2)}{\rho_{12}}\right\} - \frac{y_2}{n}\Phi\left\{\frac{\rho_{12}}{2}+\frac{\log(y_2/y_1)}{\rho_{12}}\right\} + o\left(\frac{1}{n}\right).
$$
We use (\ref{appx2-intf}) to compute other terms in (\ref{appx2-base}). Let $c_1^* = c_1-\frac{1}{\alpha_{1}^U}-\frac{\rho}{\alpha_{2}^U}$ and $c_2^* = c_2-\frac{\rho}{\alpha_{1}^U}-\frac{1}{\alpha_{2}^U}$, and let $c_{12}=\exp\left\{-\frac{c_1}{\alpha_1^U}-\frac{c_2}{\alpha_2^U}+0.5(\rho_{12}^*)^2\right\} = \exp\left(-\frac{c_1^*}{\alpha_1^U}-\frac{c_2^*}{\alpha_2^U}\right)$. We get
{\footnotesize\begin{multline*}
\int_{\mathbb{R}^1_+}I_{12}(v_0)\d v_0 = \frac{c_{12}}{n}\int_{\mathbb{R}^1}
\Phi\left(\alpha_{10}^Uv+c_1^*\,, \alpha_{20}^Uv+c_2^*\right)\exp\{-(\delta_{12}-1)v\}\d v + o\left(\frac{1}{n}\right) \\
= \frac{c_{12}}{n}\sum_{k=1}^2\alpha_{k0}^U\int_{\mathbb{R}^1}\Phi\left\{\frac{c_{3-k}^*-\rho c_k^* + (\alpha_{3-k,0}^U-\rho\alpha_{k0}^U)v}{(1-\rho^2)^{1/2}}\right\}\phi\left(\alpha_{k0}^Uv+c_k^*\right)\exp\{-(\delta_{12}-1)v\}\d v + o\left(\frac{1}{n}\right)\\
= \frac{1}{n}(\delta_{12}-1)^{-1}y_1^{\delta_1}y_2^{1-\delta_1}\delta_1^*\exp\left\{0.5\delta_1(\delta_1-1)\rho_{12}^2\right\}\Phi\left\{\rho_{12}(0.5-\delta_1)+\frac{\log(y_2/y_1)}{\rho_{12}}\right\} \\
\frac{1}{n}(\delta_{12}-1)^{-1}y_2^{\delta_2}y_1^{1-\delta_2}\delta_2^*\exp\left\{0.5\delta_2(\delta_2-1)\rho_{12}^2\right\}\Phi\left\{\rho_{12}(0.5-\delta_2)+\frac{\log(y_1/y_2)}{\rho_{12}}\right\}
 + o\left(\frac{1}{n}\right).
\end{multline*}}
Similarly, we can calculate the two remaining terms in (\ref{appx2-base}), and, after combining terms and taking limit as $n\to \infty$, we can show that (\ref{prop-1-HS}) holds.

\emph{Case 2:} $\delta_1 > 1, \delta_2 < 1$. Denote $\alpha_2^* = \alpha_2^U-\alpha_{20}^U > 0$. We get:
\begin{multline*}\int_{\mathbb{R}^1_+}I_0(v_0)\d v_0 = \int_{\mathbb{R}^1_+}\Phi_{\rho}\{c_1-\alpha_{10}^U(v_0-\log n),c_2-\alpha_{20}^U(v_0-\log n) + \alpha_2^*\log n\}\exp(-v_0)\d v_0 \\
= \frac{1}{n}\int_{-\log n}^{\infty}\Phi(c_1-\alpha_{10}^Uv)\exp(-v)\d v + o\left(\frac{1}{n}\right) = 1 - \frac{1}{n}\exp\left\{\frac{0.5}{(\alpha_{10}^U)^2}-\frac{c_1}{\alpha_{10}^U}\right\} + o\left(\frac{1}{n}\right)\\
= 1- \frac{x_1}{n}\left(1-\frac{1}{\delta_1}\right) + o\left(\frac{1}{n}\right), 
\end{multline*}
\begin{multline*}\int_{\mathbb{R}^1_+}I_{12}(v_0)\d v_0 = \frac{c_{12}}{n^{1+\alpha_2^*/\alpha_2^U}}\int_{-\infty}^{\log n}\Phi_{\rho}\left(\alpha_{10}^Uv+c_1^*\,, \alpha_{20}^Uv+c_2^*+\alpha_2^*\log n\right)\exp\{-(\delta_{12}-1)v\}\d v \\
= \frac{c_{12}}{n^{1+\alpha_2^*/\alpha_2^U}}\int_{\mathbb{R}^1}\Phi\left(\alpha_{10}^Uv+c_1^*\right)\exp\{-(\delta_{12}-1)v\}\d v + o\left(\frac{1}{n}\right) = o\left(\frac{1}{n}\right).
\end{multline*}

Similarly, we find that
$$
\int_{\mathbb{R}^2_+}I_{1}(v_0)\d v_0 = \frac{x_2}{n} + o\left(\frac{1}{n}\right)\,, \quad\int_{\mathbb{R}^2_+}I_{1}(v_0)\d v_0 = \frac{x_1}{n\delta_1} + o\left(\frac{1}{n}\right)\,.
$$
Therefore, $\ell(x_1,x_2) = x_1 + x_2$. This implies that $C_{2,1:2}^{\Ww}$ as well as the limiting extreme-value copula $\mathcal{C}_{2,1:2}^{\Ww}$ has no upper tail dependence. The remaining two cases when $\delta_1 < 1, \delta_2 > 1$ and when $\delta_1 < 1, \delta_2 < 1$ are considered analogously. $\hfill \Box$

\end{document}